# Machine Learned Hückel Theory: Interfacing Physics and Deep Neural Networks

Tetiana Zubatyuk[1]*, Ben Nebgen[2]*, Nicholas Lubbers[2], Justin S. Smith[2], Roman Zubatyuk[1], Guoqing Zhou[3], Christopher Koh[2], Kipton Barros[2], Olexandr Isayev[1], Sergei Tretiak[2]

*Contributed Equally

Affiliations:

1. UNC Eshelman School of Pharmacy, University of North Carolina at Chapel Hill, Chapel Hill, NC 27599, USA.
2. Los Alamos National Laboratory, Los Alamos, NM 87544, USA.
3. University of Southern California, Los Angeles, CA 90089, USA.

## Abstract

The Hückel Hamiltonian is an incredibly simple tight-binding model famed for its ability to capture qualitative physics phenomena arising from electron interactions in molecules and materials. Part of its simplicity arises from using only two types of empirically fit physics-motivated parameters: the first describes the orbital energies on each atom and the second describes electronic interactions and bonding between atoms. By replacing these traditionally static parameters with dynamically predicted values, we vastly increase the accuracy of the extended Hückel model. The dynamic values are generated with a deep neural network, which is trained to reproduce orbital energies and densities derived from density functional theory. The resulting model retains interpretability while the deep neural network parameterization is smooth, accurate, and reproduces insightful features of the original static parameterization. Finally, we demonstrate that the Hückel model, and not the deep neural network, is responsible for capturing intricate orbital interactions in two molecular case studies. Overall, this work shows the promise of utilizing machine learning to formulate simple, accurate, and dynamically parameterized physics models.

## I. INTRODUCTION

An accurate understanding of interacting electrons and nuclei is required in materials and chemistry research[1]. Quantum mechanical (QM) many-body calculations for solving the Schrödinger equation (SE) provide a generally accurate and reliable method for obtaining descriptions of such interactions. Development of efficient approximate methods for solving the SE remains a grand challenge.[2] Such approximate methods are in high demand by the fields of molecular biology, materials science and pharmaceutical science, which require the consideration of extended and complex systems. It is well known that the computational cost of solving the SE through the variational procedure grows exponentially with the number of electrons. The most accurate methods (e.g. configuration interaction (CI) or coupled-cluster (CC) methods) hit an exponential wall with only 10s of atoms, making such calculations intractable. For large molecular systems ($10^2$-$10^3$ atoms) the explicit calculation of the many-electron wavefunction can be replaced by density functional theory (DFT)[3,4], which scales cubically ($O(N^3)$) with the number of orbitals. The DFT single-particle energies in practice give a reasonable approximation to the actual electronic structure of a molecule. However, self-consistent DFT calculations are still too computationally demanding for some applications, such as the computation of properties from molecules with more than $10^4$ atoms or millions of independent molecules for high throughput screening applications.

A common approach that scales better for large molecular systems is to use semi-empirical effective Hamiltonians, such as the tight-binding formulation[5–8]. In particular, the Hückel method, originally designed for π-orbitals, established important chemical insights to conjugated hydrocarbon systems. The method was generalized to the Extended Hückel Method (EHM), with landmark work by Helmholz[9] and Hoffman[10] in the context of generic organic and inorganic systems. Despite some limitations[11,12], numerous studies have employed the Extended Hückel framework as a compromise between accuracy and speed that allows for inspection of quantum phenomena at a qualitative level[13–21]. However, due to the simplicity of the extended Hückel Hamiltonian, constant parameterizations will never result in a generally applicable and quantitatively accurate model. While others have explored the possibility of extended Hückel parameterization schemes that depend on bond distance, accuracies in the kcal/mol range will require parameterizations that are more complicated functions of the atomic environment[22]. This need calls for automated parameterization algorithms using modern data science approaches.

In the last decade, machine learning methods have become increasingly popular for predicting properties obtained by solutions to the Schrödinger equation, e.g. potential energy[23–28], forces[29–31], electronic multipole moments[32], atomic charges[33,34] or even multiple of these properties[35,36]. These methods aim to directly predict properties with linear computational scaling, bypassing the need for solving the Schrödinger equation. However, since these methods avoid the Schrödinger equation they lack any quantum description of the molecular physics, preventing inspection of the electronic structure of a given system. Providing a quantum description of molecular physics provides a richer understanding of model predictions, and is an emerging avenue of research. For example, recent research from Li et al. uses a neural network model to predict the parameters of a density functional TB Hamiltonian for a given set of atomic coordinates, which allows accurate prediction of reference QM data[37]. Welborn et al. deployed a machine learning model to predict the correlation energy computed by accurate yet expensive post-Hartree-Fock methods using less expensive Hartree-Fock intermediate quantities as descriptors[38].

In this article, we introduce a machine learning-based semi-empirical TB scheme based-on the extended Hückel model. The key idea of the scheme is to learn the matrix elements of a one-electron Hamiltonian as a function of local atomic environments. To the best of our knowledge, this is the first study of ML models used within the context of extended Hückel theory. Using the HIP-NN[25] neural network architecture, we predict a Hückel model Hamiltonian that is responsive to a specific molecular geometry. Our model, dubbed EHM-ML, accurately reproduces DFT electronic structure calculations with many orders of magnitude speedup over DFT. It generates transferable parameters and provides a computational tool for carrying out calculations on large molecular systems. Moreover, we show via case studies that the wavefunctions of our EHM-ML Hamiltonian reflect the same physical processes that take place in DFT reference calculations. The key advance in this work is the demonstration that the effective Hamiltonian, and not the ML model, performs most of the work associated with capturing the orbital physics observed in these molecular systems.

We detail the EHM-ML model architecture and training procedure in Section 2. In Section 3, we demonstrate that HIP-NN effectively learns to parameterize extended Hückel theory and test the EHM-ML model on the COMP6[39] benchmark of large organic molecules. In Section 4, the inner workings of the EHM-ML model are explored in a pair of case studies. The first study discusses the implementation of our EHM-ML Hamiltonian on a methane molecule and explores the behavior of the EHM parameterization over a bond vibration. The second case study examines internal rotation around single bonds in butadiene and aza-butadiene. Finally, we conclude in Section 5.

## II. EHM-ML FRAMEWORK
### 2.1 the Extended Hückel model

Like all Hückel method [40–42] counterparts, the EHM calculations employ a simplistic tight-binding form of the Hamiltonian, which is taken as a sum of single-electron interactions. Linear Combinations of Atomic Orbitals (LCAOs) ($|\varphi\rangle = \sum_{\mu=1}^{N} c_\mu |\chi_\mu\rangle$) are further used to represent molecular orbitals (MOs) without an anti-symmetrization operator, where the basis set is limited to the valence AOs of the constituent chemical elements. Traditionally, Hückel theory solves the Schrödinger's equation in a non-orthonormal basis, thus requiring a generalized eigensolver algorithm to address the following matrix diagonalization problem, where $S$ is an overlap matrix between the basis functions $|\chi_\mu\rangle$.

$$H\Psi = ES\Psi \quad (1)$$

Specifically, the definition of the Hückel theory Hamiltonian $H$ is given by

$$H = \begin{bmatrix} \alpha_{11} & \frac{1}{2}K^{\ddagger}(\alpha_{22}+\alpha_{11})S_{21} & \cdots & \frac{1}{2}K^{\ddagger}(\alpha_{NN}+\alpha_{11})S_{N1} \\ \frac{1}{2}K^{\ddagger}(\alpha_{11}+\alpha_{22})S_{12} & \alpha_{22} & \cdots & \frac{1}{2}K^{\ddagger}(\alpha_{NN}+\alpha_{22})S_{N2} \\ \vdots & \vdots & \ddots & \vdots \\ \frac{1}{2}K^{\ddagger}(\alpha_{11}+\alpha_{NN})S_{1N} & \frac{1}{2}K^{\ddagger}(\alpha_{22}+\alpha_{NN})S_{2N} & \cdots & \alpha_{NN} \end{bmatrix} \quad (2)$$

where $N$ is the total number of basis functions and $H_{\mu\mu} = \alpha_{\mu\mu}$ is an orbital energy. For the remainder of the paper $\mu$ and $\nu$ will refer to indices over Hamiltonian elements while $I$ and $J$ will refer to atomic indices. The off-diagonal elements $H_{\mu\nu}$ are computed with the Wolfsberg-Helmholtz formula[9]

$$H_{\mu\nu} = \langle \chi_\mu | \widehat{H} | \chi_\nu \rangle = \beta_{\mu\nu} = \frac{1}{2}K^{\ddagger}(H_{\mu\mu} + H_{\nu\nu})S_{\mu\nu} \quad (3)$$

where $K^{\ddagger}$ is an empirically fit parameter. In the construction of the Hamiltonian, $S$ is not assumed to be the identity and the overlap integrals are computed, for example, using the Slater form of the valence basis functions $|\chi_\mu\rangle$ as is done in the YAeHMOP[43] code (extended Hückel calculation and visualization package).

***Choice of*** $\boldsymbol{\alpha_{\mu\mu}}$. Traditionally, the diagonal $\alpha_{\mu\mu}$ energy parameters in the EHM method are the negative of the ionization energies (IE) of valence state for the appropriate isolated atom, as given by Koopmann's theorem[44]. However, this approximation neglects the fact that an atom may have several electronic valence states. For example, in the isolated carbon atom, the lowest-energy states are associated with the configuration $1s^22s^22p^2$. In a saturated molecule such as methane, carbon shares electrons with four hydrogens and behaves as though it were in the $2sp^3$ configuration. Various authors recommend slightly different sets of valence IEs [45,46]. Furthermore, the proper valence state for carbon in methane differs from that in ethylene, which in turn differs from that in acetylene. Generally, this is ignored in the EHM approach and a compromise set of valence state IEs is selected for use over an entire range of molecules.

***Choice of*** $K^{\ddagger}$. The coefficient $K^{\ddagger}$ is an additional dimensionless empirical fitting parameter which depends on the state of matter being studied. [47] For molecules, the EHM model was found to agree with experiment best when $K^{\ddagger} = 1.75$, while for solids a value of 2.3 is optimal. Despite $K^{\ddagger}$ being constant, the inclusion of overlaps in off diagonal elements ($H_{\mu\nu}$) of the Hamiltonian enforce that well separated orbitals do not directly interact.

## 2.2 Improvement of EHM with Machine Learning

Interfacing ML with the above Hamiltonian is accomplished by replacing the static parameters $\alpha_{\mu\mu}$ and $K^{\ddagger}$ with their ML predicted counterparts. For instance, due to the dependence of $\alpha_{\mu\mu}$ on the valence state of an atom as discussed in Section 2.1, it would be desirable to have an orbital energy parameter depend on the atomic environment. Having non-constant Hückel parameters is not a new idea. For example, Calzaferri and coworkers [19,48] suggested corrections to coefficient K; thereafter the proposed weighted distance-dependent expression was widely used.[22] However, distance alone is an insufficient modulator for $K^{\ddagger}$ as the properties of a bond depend substantially upon higher-order features of the entire chemical environment of each atom, e.g. bond order, conjugation, and hybridization. It would be nearly impossible to imagine all features of the chemical environment contributing to a bond parameter and simplify them into a mathematical expression for $K^{\ddagger}$. Fortunately, determining such an unknown functional form is the forte of a deep neural network.

ML frameworks for quantum chemistry attempt to capture the chemical environment[24,25,49,50] around an atom and use this information to assign a value (frequently atomic contribution to total energy) to each atom. Instead of using a pre-defined functional form to make this value assignment, the deep neural network is allowed to learn the implicit functional form, subject to a few constraints such as continuity and differentiability. Thus, instead of trying to develop a functional form for $\alpha_{\mu\mu}$ and $K^{\ddagger}$ manually, a deep neural network will determine these parameters automatically, taking into account the current chemical.

### 2.3 HIP-NN implementation of EHM-ML

Figure 1 illustrates the general scheme of the present EHM-ML model used for predicting electronic properties of molecules. The input (Fig 1, left) is the molecular configuration, which uses a simple representation that is symmetric with respect to translation, rotation, and permutation of atoms. The Hierarchically Interacting Particle Neural Network (HIP-NN) generates the set of parameters $\{\alpha_{ii}, K^{\ddagger}\}$ (blue boxes), accounting for the complete chemical environment of each atom and atom pair, to create an effective extended Hückel Hamiltonian. It is important to emphasize that, in contrast to the original EHM formulation, the parameters $\{\alpha_i, K^{\ddagger}\}_i^k$ are continuously varying with molecular geometry. HIP-NN is a message passing neural network written in Theano and Lasangna[51,52] designed to make accurate predictions of quantum mechanical properties on molecules.[25] Green boxes denote interaction layers, which mix information between atoms. Red boxes denote on-site layers, which process the atomic features of a single atom. A detailed description of HIP-NN can be found in the original paper.[25]

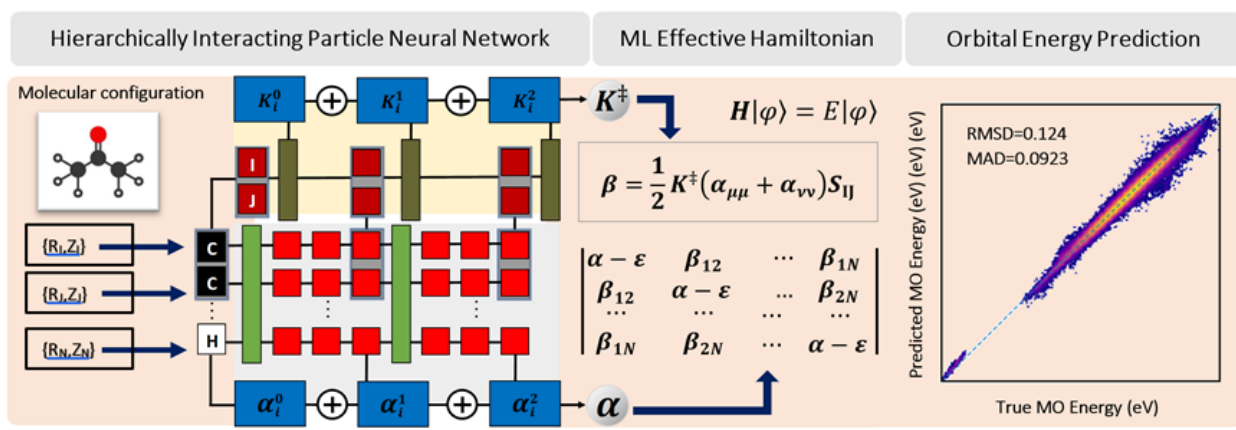

Figure 1: Scheme of the EHM-ML model. HIP-NN uses the atomic coordinates $R_I$ and atomic numbers $Z_I$ to generate custom EHM Hamiltonian matrix elements, $\{\alpha, \beta\}$.

Due to limitations in the Theano framework, a generalized eigensolver was not avalable.[51] Therefore, in the matrix diagonalization step, and only in this step, we assume that $S$ is the identity matrix. This approximation is not unprecedented as there are a large body of neglect of differential overlap (NDDO) effective Hamiltonian methods, such as AM1 and PM3, that successfully utilize the same appoximation.[53–55] Additionally, the learned parameters have the ability to account for $S$, since left multiplying Eq. 1 by $S^{-1}$ shows that the elements of the overlap matrix can be whole numbers combined with the learned $\{\alpha_{ii}, K^{\ddagger}\}$.

## 2.4 Augmentation of HIP-NN for $K^{\ddagger}$

HIP-NN has been used to predict atomic energies,[25] charges,[34] and dipoles[32] in terms of local atomic charge or energy approximations. To predict $K^{\ddagger}$, a quantity that describes the interaction of two atoms in the off-diagonal elements in Eq. 1, we introduce a new type of layer into the HIP-NN framework. The *pair regression layer* produces outputs for each pair of atoms in the system. At every hierarchical level $\ell$, there is a learnable matrix as a function of distance $u^{\ell}_{ab}(r)$, where the $a$ and $b$ indices contract over the pair of atomic features:

$$u^{\ell}_{ab}(r) = \sum_v U^{\ell}_{\tau,ab} s^{\ell}_{\tau}(r) \tag{4}$$

The sensitivity functions $s$ are parameterized as in the Interaction Layers (Ref 25, eq. 10), and $\tau$ indexes the sensitivity functions. $U^{\ell}_{v,ab}$ is a three dimentional tensor of learnable parameters. The pair-valued $K^{\ddagger\ell}_{IJ}$ is computed as a bilinear function in the atomic features $z$, symmetrized to impose the Hermitivity of the Hamiltonian:

$$K^{\ddagger\ell}_{IJ} = \sum_a \sum_b z^{\ell}_{I,a} u^{\ell}_{ab}(r_{IJ}) z^{\ell}_{J,b} + (I \leftrightarrow J) \tag{5}$$

Here, $a$ and $b$ index feature vectors on atoms $I$ and $J$ respectively. The prediction for $K^{\ddagger}_{IJ}$ is constructed by summing over the contributions from each hierarchical block $\ell$ :

$$K^{\ddagger}_{IJ} = \sum_{\ell} K^{\ddagger\ell}_{IJ} \tag{6}$$

The predicted $K^{\ddagger}_{IJ}$ matrix exists over atom pairs; whereas the $K^{\ddagger}_{\mu\nu}$ needed for EHM exists over orbital pairs. To resolve this dimension mismatch, the predicted $K^{\ddagger}_{IJ}$ is expanded over the orbital indices such that all orbital pairs that share atom centers also share the same $K^{\ddagger}_{\mu\nu}$ parameter. The learning algorithms are set up to reproduce the DFT reference orbital energies and respective densities for organic molecules containing four atom types: H, C, N, and O, as outlined in the next subsection. HIP-NN utilizes a DFT labeled training dataset to automatically learn representations of local atomic environments without requiring any feature engineering [25].

### 2.5 Loss function for training

We train the EHM-ML model by varying the neural network parameters to minimize a loss which quantifies the error of a model in performing a task. Since our goal with the EHM-ML model is to learn quantum mechanics-based molecular physics, we train it to reproduce properties of DFT calculated molecular orbitals. We start by including an error term corresponding to the molecular orbital energy spectrum. However, our DFT data and EHM model are defined over different basis sets. In order to define a matching between the two sets of orbitals, we examine the molecular orbitals surrounding the occupation threshold. We take X occupied orbitals and Y virtual orbitals, order them by energy, and compute the difference between EHM-ML and DFT squared:

$$Orbital\ Energy\ Error\ (OEE) = \sum_{j=HOMO-X+1}^{LUMO+Y-1} \left(\varepsilon_j^{DFT} - \varepsilon_j^{Hückel}\right)^2 \quad (7)$$

where index $j$ labels the orbitals. However, such a loss term does not fully constrain the EHM-ML model; a Hamiltonian may reproduce the eigenspectrum of DFT without reproducing the distribution of the electronic density over MOs. This is particularly important when preserving the orbital identities when their energies cross depending on molecular conformation. We address the orbital density distribution by defining an *Orbital Occupation Fractions (OOF)* for orbital $\rho$ on atom $I$:

$$q_{j,I} = \frac{\Sigma_{\mu \in \{I\}} C_{j,\mu}^2}{\Sigma_\mu C_{j,\mu}^2} \quad (8)$$

The notation $\mu \in \{I\}$ denotes that the sum in the numerator runs over all atomic orbitals associated with atom I; the sum in the denominator runs over all atomic orbitals in the molecule. This definition provides positive-definite atom-local charge densities for each molecular orbital, which can be compared between basis sets of different sizes. The sum-squared error of the OOF is the *Density Error*:

$$Density\ Error\ (DE) = \sum_j^{N_{orb}} \sum_I^{N_{atom}} \left(q_{j,I}^{DFT} - q_{j,I}^{Hückel}\right)^2 \quad (9)$$

The full loss function $\mathcal{L}$ is a combination of OEE, DE, and an L2 regularization penalty using importance factors λ:

$$\mathcal{L} = \lambda_{OEE} \times OEE + \lambda_{DE} \times DE + \lambda_{L2} \times \mathcal{L}_{L2} \qquad (10)$$

The $\mathcal{L}_{L2}$ regularization penalizes large weight values and promotes smoothness of the network. We implement this with eq. 16 of ref 25. λ values were chosen empirically to be $\lambda_{OEE}=1.0$, $\lambda_{DE}=10^{-3}$, $\lambda_{L2}=10^{-6}$. Interestingly, the inclusion of the DE term during training actually results in better minimization of the OEE term when compared to a model trained without DE. We take this as evidence that the inclusion of DE successfully promotes the construction of physically meaningful Hamiltonians that correctly track the character of MOs.

## 2.6 Training Data Set

To train the HIP-NN neural network model to predict the EHM-ML model parameters, we use a small fraction of the ANI-1x dataset[39] of non-equilibrium small organic molecule conformations containing the elements C, H, N, and O. The ANI-1x dataset was previously generated in an active learning process with ANI neural network model potential[24]. The final dataset contains approximately 5 million DFT (wB97x/6-31g*) calculations, each computed with a neutral charge state and singlet spin state. These quantum calculations, along with all other quantum calculations done in this work were performed with Gaussian09.[56] All molecules in this data set are obtained from a variety of sources: GDB-11[57,58], ChEMBL[59], and tripeptides generated from RDKit cheminformatic software package[60]. Multiple methods for sampling are used in non-equilibrium sampling for molecules in the data set, including normal mode sampling, molecular dynamics sampling, and molecular dynamics dimer sampling. Details about the dataset can be found in our previous publication.[39]

A fundamental limitation of the model is that it cannot back propagate through a system with degenerate orbitals, due to the computation of the backward gradient involving the difference of eigenvalues in the denominator. Therefore, this model is unable to train on molecules with symmetry enforced degenerate orbitals, though prediction is not effected. In practice, degenerate orbitals only occur in ANI-1x in atomic dimers, where the two rotationally symmetric p-orbitals are perfectly degenerate. This is because ANI-1x was generated through random vibrational perturbations from equilibrium structures and molecules with symmetry are very sparse in the space of all molecules. For this reason, all dimer molecules were excluded from ANI-1x before training or testing, though in principal the model could predict on dimer molecules.

Once the atomic dimers were trimmed, all molecules from ANI-1x with five atoms or less were forced to be included in the training set. This was found to aid ML in producing realistic parameters for the EH model. In addition to these molcules, a random sample of the remaining molecules with 18 atoms or fewer was added to the training set with no biases on the size or type of molecule. 18 atoms were used as a cutoff for the training set due to poor scaling of the training code to larger system sizes, since overlap matrices were zero-padded to the largest possible Hamiltonian size. Three distinct models were trained, each to a different number of occupied and virtual orbitals. The final sizes of the training datasets and target orbitals are given in Table 1. Section 1 of the SI contains histograms of the sizes of molecules found in all of the datasets used in this paper.

Table 1: Training set sizes for the three models explored in this paper.

| Model | Training Orbitals | Molecules in Training Set |
|---|---|---|
| 4-Occ 2-Virt | HOMO-3 to LUMO+1 | 172,588 |
| 4-Occ 0-Virt | HOMO-3 to HOMO | 210,933 |
| 2-Occ 2-Virt | HOMO-1 to LUMO+1 | 210,652 |

## III. RESULTS AND DISCUSSION
### 3.1 Test set accuracy

Fig. 2 compares the accuracy of three EHM-ML models with different orbital configurations. The comparison is performed by predicting on the remainder of ANI-1x (described in Sec. 2.6) with 20 atoms or fewer. The size limit was imposed so that we would have a measure of how well the models perform on molecules similar to those in the training set. Perhaps unsurprisingly, the "4-Occ 0-Virt" model has the best predictive performance on the remaining ANI-1x data. It can predict the energies of the four highest occupied orbitals in training set with a MAD of 0.092eV. This model is completely unable to predict the band gap of organic molecules (fig. 2h) which is anticipated since it was never trained to virtual orbitals. Models trained to virtual orbitals ("4-Occ 2-Virt" and "2-Occ 2-Virt") show a lower accuracy when predicting orbital energies (fig. 2a,c). This is likely due to the EHM lacking explicit coulomb and exchange terms, making the model insufficiently complex to describe both the occupied and virtual orbitals of Density Functional Theory. Despite this, the "4-Occ 2-Virt" model can predict band gaps to about 0.2 eV (fig. 2i). None of the three models have true quantitative accuracy when predicting orbital densities. This is likely due to the reduced importance in the loss function for orbital densities. It is still important to note that there is roughly an order of magnitude more points near the diagonal of the orbital density correlation plots than there are near the edges. This indicates that the EHM-ML model learned some properties of the orbital shapes qualitatively.

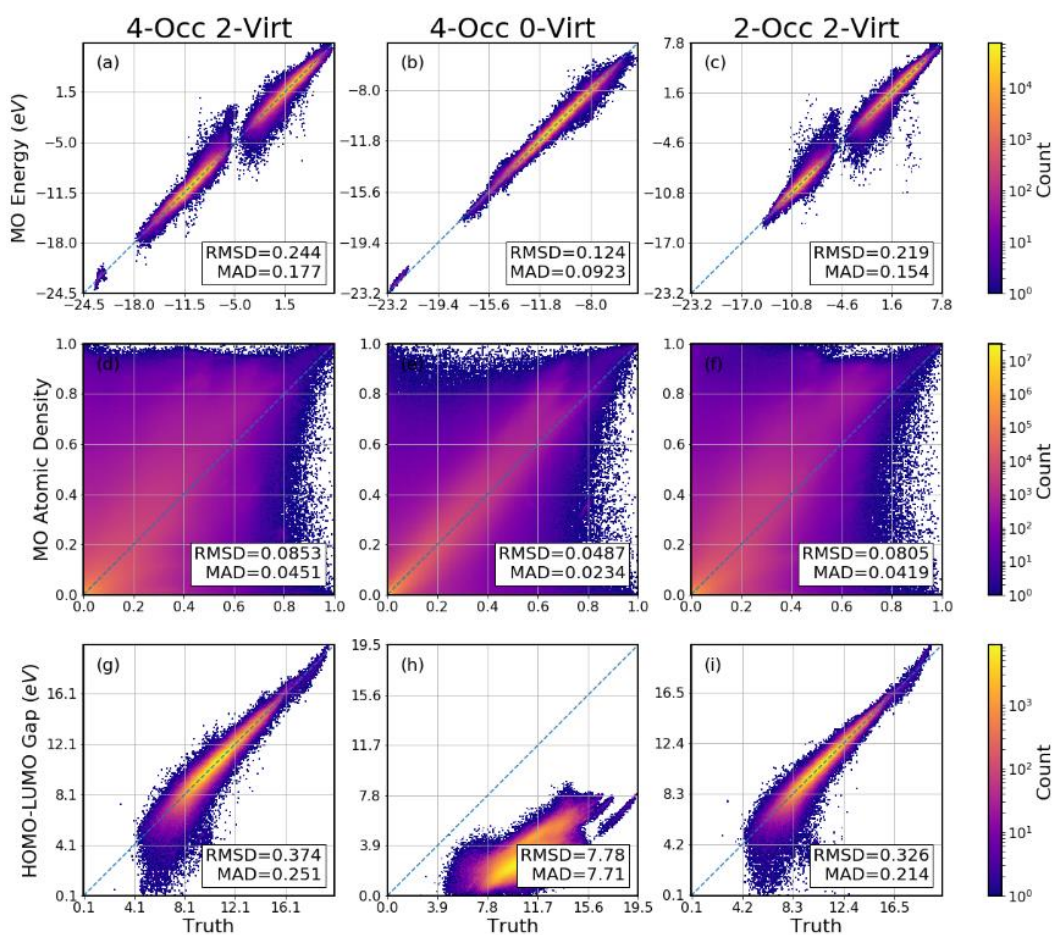

Figure 2: Predictions of EHM-ML on the remaining ANI-1x data that were not included in the training sets. This includes some large molecules than were intentionally left out of training. While the 4-Occ 0-Virt model makes excellent occupied orbital predictions (b), it is entirely unable to predict the band gap (h). Other models trained to virtual orbitals give worse predictions on orbital energies (a,b), but are able to make predictions on the band gap (g,i). While all of the models have outliers in the orbital density predictions (d,i,f), there are are order of magnitude more predictions near the diagonal than there are qualitatively wrong predictions.

## 3.2 Validation on COMP-6

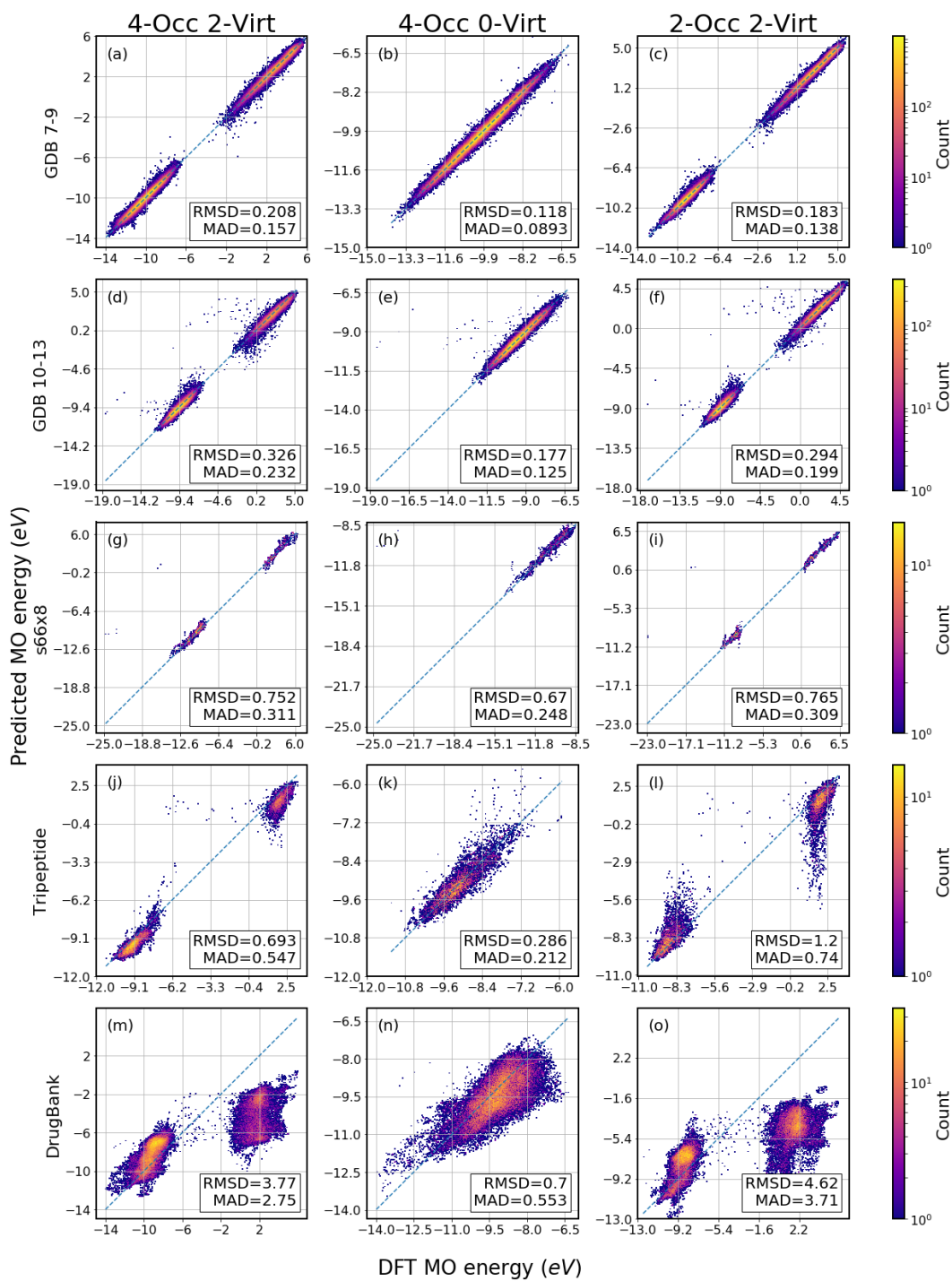

Figure 3 Prediction of the EHM-ML models on the COMP-6 test set. Generally, the models perform well on all subsets except drugbank. Even tripeptides, which contain molecules up to 69 atoms in size, is well reproduced (j,k,l).

Fig. 3 shows the performance of our EHM-ML model on a portion of the COMP6 benchmark[39] to validate the model's ability to extend its predictions to molecular systems larger than those in the training dataset. The portion tested consists of five separate benchmarks: GDB 7-9, GDB 10-13, s66x8, Tripeptide, and DrugBank. The GDB benchmarks are randomly subsampled molecules from the GDB-13 benchmark containing the number of heavy atoms within the range given in the test sets name. Each benchmark contains a few thousand different molecules. Normal mode sampling is used to generate tens of thousands random conformations from those chosen molecules. The s66x8 benchmark[61] is an existing benchmark for measuring the interaction of dimer molecules. The tripeptide benchmark contains 248 3-amino acid peptide chains, which were randomly generated using the RD-Kit chemiformatics package. Normal sampling was used to generate approximately 2,000 non-equilibrium conformations. The DrugBank benchmark was built by randomly subsampling the DrugBank database[62] then carrying out normal mode sampling to generate 13,000 conformations. All test sets are described in more detail in the previous work[39].

Generally, the performance of our EHM-ML model is very good. Orbital predictions on the subsets that are most similar to ANI-1x, GDB 7-9 and GDB 10-13, by the "4-Occ 0-Virt" have mean absolute deviations of 0.089 and 0.13 eV, respectively. Additionally, while performance degrades by roughly a factor of 2 when applying the model to the much larger systems found in tripeptides. Another interesting comparison is between the "4-Occ 2-Virt" and "2-Occ 2-Virt" networks. While the "2-Occ 2-Virt" network performs slightly better on the GDB datasets, which are similar to the training set, the "4-Occ 2-Virt" does significantly better on the tripeptides dataset. This is especially pronounced in the band gap predictions, which can be found in Section 3 of the SI, where the "4-Occ 2-Virt" has errors 3 times smaller than the "2-Occ 2-Virt" model. This is likely due to the "2-Occ 2-Virt" model having fewer constraints during fitting, allowing it to be more accurate on similar data. However, the additional physics provided by the extra occupied orbitals allows the "4-Occ 2-Virt" model to extend to larger systems better. Finally, the poor performance of all networks on the DrugBank dataset is likely due to the lack of aromatic rings in the training data. The structures producing the highest error in the s66x8, tripeptides, and DrugBank datasets, of which nearly all contain ring structures, can be seen in Section 4 of the SI.

Fig. 4 shows histograms of the parameters generated by the "4-Occ 0-Virt" model when predicting on COMP-6. These are the $\alpha_{ii}$ values for s-orbitals fed into the EHM model. One interesting point of agreement between the neural network and the original Hückel theory is the prediction of the hydrogen s-orbital energy. In extended Hückel theory, this parameter was taken from the ionization energy of Hydrogen, which is 13.6 eV. By training the neural network on molecular orbital energies, the "4-Occ 0-Virt" network produces an average hydrogen s-orbital energy of 12.5 eV, which is very close to the extended Hückel theory value. Another interesting observation is that the sequence of orbital energies (O<N<C<H) is in line with the expected sequence from simple electrostatic arguments. While not shown, the predictions of $K^{\ddagger}$ are also in line with the original EHM parameterization, which HIP-NN producing a bifurcated distribution with an average value of 1.54, which is similar to the original Hückel parameterization of 1.75. Figures detailing predictions made by the other networks as well as the distribution of $K^{\ddagger}$ are shown in Section 2 of the SI.

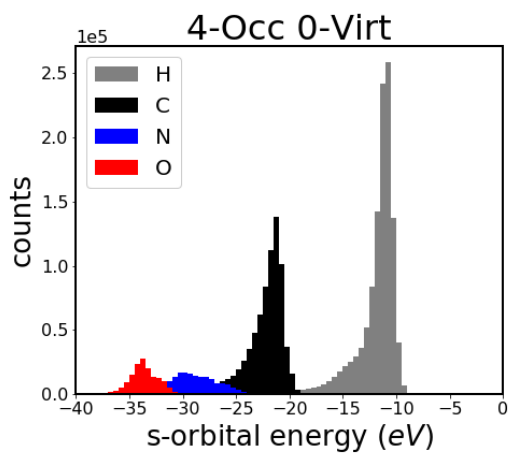

Figure 4: Histogram of ML predicted Hückel theory parameters for the "4-Occ 0-Virt" model over the COMP-6 test set. The hydrogen s-orbital energy (a,b,c) is in remarkable agreement with the original Hückel theory assignment of -13.6 eV even though the ML model was given no artificial bias to this value. Additionally, the ordering of the orbital energies follows electrostatic arguments, with carbon having the highest atomic orbital energy of the heavier elements having the lowest.

## IV. REPRESENTATIVE MOLECULAR CASE STUDIES
### 4.1 Molecular Orbital View of C-H stretching in methane

The orbital shapes and energies of valence electrons are of crucial importance to obtaining a wide variety of quantum mechanical properties. Example uses of these orbital shapes and energies are Kenichi Fukui's frontier electron orbital theory of reactivity[63] and the Woodward-Hoffmann rules[64], where information on orbital distributions and phase is required (not just total electron density). Effective Hamiltonian models that correctly describe the behavior of valence electrons are required for such applications. In this section, we examine the accuracy of our EHM-ML model on describing the orbital structure of a C-H stretch of a methane molecule.

The "4-Occ 0-Virt" model was employed to find the orbital energies and wavefunctions of the four highest energy molecular orbitals of methane. In testing, we found that only this model was able to capture the interesting orbital physics detailed below. This is likely due to the simplicity of the Hückel Hamiltonian; it is unable to make simultaneous quantitative predictions on occupied and virtual orbitals. In Fig. 5a we plot the EHM-ML predicted orbital energies $\alpha_{\mu\mu}$ along with the original parametrization used by Pople and Segal[46]. The original approach uses static parameters, whereas our EHM-ML model reacts smoothly to the bond stretch. The dissociating hydrogen ($H^a$) 1S orbital increases as the hydrogen is extracted. As a result of increasing site energy, one would expect the hydrogen to decouple from the lowest energy molecular orbitals, reflecting a dissociation of bonding between the hydrogen and the $CH_3$ fragment. At the same time the carbon 2S orbital energy decreases by a similar amount, while the carbon 2P orbital energies are approximately constant. Near the equilibrium bond distance (~1.09 Å for our reference DFT), all on-site energies are roughly static. This provides strong evidence that static formulations of Extended Hückel Theory are only appropriate near equilibrium geometries.

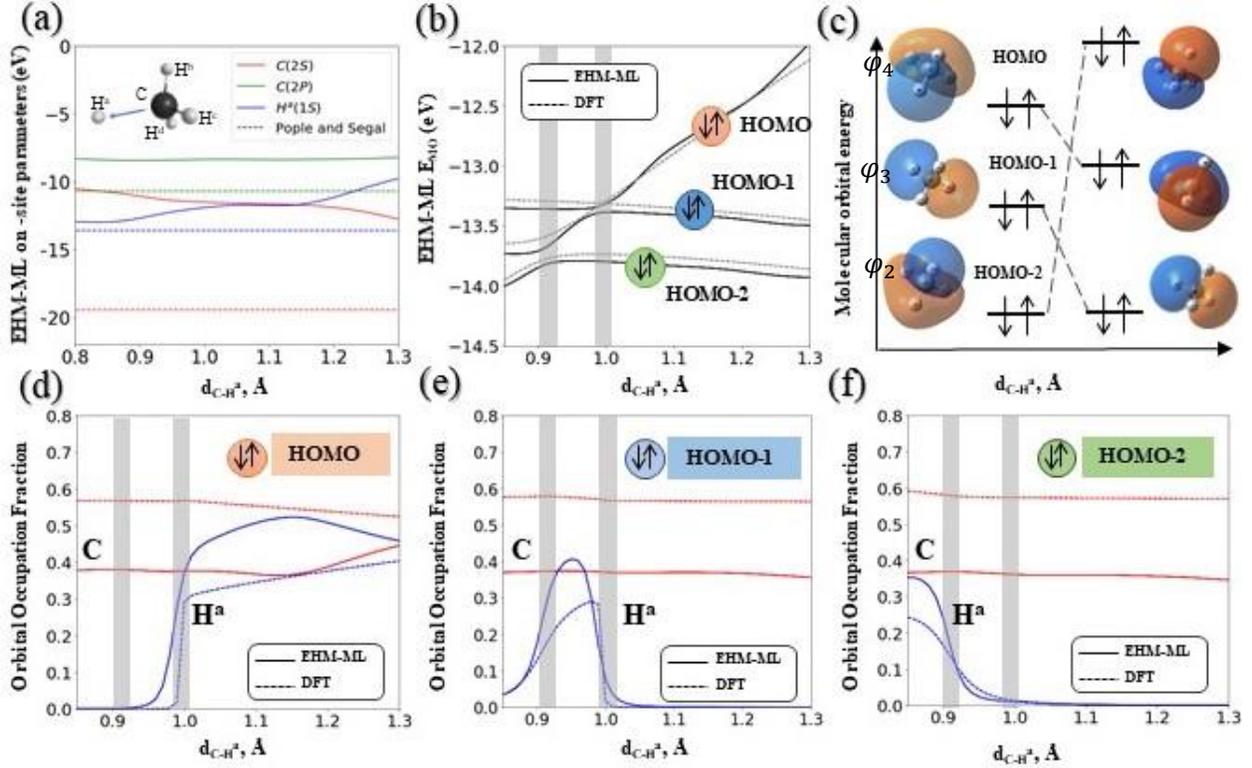

Figure 5: EHM-ML parameterization (a) and orbital energies (b) as a function of hydrogen extraction coordinate. Partial MO diagram (c) which illustrates how the frontier MO (FMO) change in the process of C-H$^a$ bond stretching. The two grey areas on the diagrams indicate the FMO crossings between HOMO-2 and HOMO-1, HOMO and HOMO-1. Frames d-f depict the OOF on carbon and extracted hydrogen for the highest lying frontier orbitals.

In Fig. 5 we show that the dynamic EHM-ML parameters correctly reproduce the changes in electronic structure witnessed in the source DFT throughout the hydrogen extraction coordinate. This is shown through MO energies and charge occupation via the OOF $q_{\rho,I}$ (Eq. 8). Critically, Fig. 5b indicates that both DFT and EHM-ML produce two orbital crossings throughout the methane dissociation coordinate: one at ≈0.9 Å between HOMO-2 and HOMO-1 and one at ≈1.0 Å between HOMO-1 and HOMO. By plotting the DFT orbitals (Fig. 5c), it is clear that the HOMO-2 orbital at a C-H distance of 0.85 Å becomes the HOMO orbital when the C-H distance is 1.15 Å. Meanwhile, the HOMO and HOMO-1 orbitals at a C-H distance of 0.85 Å are shifted downward but maintain their order as the C-H bond is extended.

The correspondence between orbital energy and orbital character as determined by the OOF is of particular interest. As the HOMO-2 and HOMO-1 orbitals approach in energy near a C-H separation of ≈0.9 Å, the character of these two orbitals switch as indicated by Fig. 5e, and f. The HOMO-1 $H^a$ OOF starts near zero and transitions to 0.5 $e^-$. The opposite is observed for the HOMO-2 $H^a$ OOF, indicating that around 0.9 Å these two orbitals switch character. At 1.0 Å, another closer crossing between the HOMO and HOMO-1 orbitals appears. Correspondingly, a more abrupt exchange of orbital character, as indicated by the $H^a$ OOFs, is shown in Fig. 6d and f near 1.0 Å. Both the gentle crossing at ≈0.9 Å and the abrupt crossing at ≈1.0 Å are captured by the EHM-ML model without rapid changes in EHM parameters, as shown in Fig. 5a. This example illustrates how the NN and EHM pieces of EHM-ML can harmonize to capture orbital physics across large changes in geometry; EHM can capture rapid changes in orbital physics using diagonalization, and the NN parameters provide soft modulation of the underlying Hamiltonian.

### 4.2 Internal rotation in conjugated systems

The key success of the original Hückel model was the predictions it made on the behavior of conjugated π systems in electrocyclic reactions. This became the theoretical background for the Woodward-Hoffman rules and the subject of the 1981 Nobel Prize in Chemistry. While limitations in the training dataset preclude the application of our model to reactive systems, it can be applied to bond rotations in conjugate π systems. To illustrate this we apply our EHM-ML model to the rotation about the center bond in 1,3-butadiene and 2-aza-1,3-butadiene. While these two systems have the same number of electrons, aza-butadiene has one fewer core s-orbital. Thus, the LUMO in butadiene becomes the HOMO in aza-butadiene with a similar shift happening between all valence orbitals. This can be clearly seen when examining the π system on the right side of Fig. 6. We have already checked that a change in EHM-ML energy of FMO reflects certain changes in calculated OOF populations of corresponded AOs. We now consider the internal rotation in conjugated systems to check that such an EHM-ML FMO energy change agrees directly with the structural changes of the molecular system.

***Orbital correlation between s-trans/s-cis-1,3-butadiene.*** In this section the s-cis-1,3-butadiene and cis-2-aza-1,3-butadiene conformers with a torsion angle Θ rotation from 180° to 0° is considered. The ground electronic states of s-trans- and s-cis-1,3-butadiene and 2-aza-1,3-butadiene conformers have closed shells with 15 doubly occupied orbitals. There are four core orbitals and four inner valence orbitals; the remaining seven orbitals are outer valence orbitals. Here we will apply our "4-Occ 0-Virt" EHM-ML model to predict the behavior of the 4 highest lying occupied orbitals in both butadiene and aza-butadiene as they undergo a torsional bond rotation. In cis-butadiene, the 4 highest orbitals have the following symmetry:

$X^1A_1$: **(6b$_2$)(7a$_1$)(1b$_1$)(1a$_2$)**

While in trans-butadiene, the symmetries of the highest orbitals are:

$X^1A_g$: **(6b$_u$)(7a$_g$)(1a$_u$)(1b$_g$)**

Rotation of the torsion causes the butadiene to lose a mirror plane and gain an inversion center, causing the point group to change from $C_{2h}$ to $C_{2v}$. Additionally, when the bond rotates, orbitals that were symmetric under the mirror plane will transform into orbitals that are symmetric under inversion, with the same holding true for orbitals that are anti-symmetric under reflection and inversion. This simple symmetry argument predicts a crossing between the HOMO and HOMO-1 in butadiene as the molecule undergoes a torsional rotation. As seen in Fig. 6, both DFT and EHL-ML find that the HOMO and HOMO-1 orbitals cross when butadiene undergoes a torsional rotation, and the EHL-ML model remains in quantitative agreement with the energy levels of DFT. For an analysis of the on-site orbital parameters constructed by HIP-NN, which change very little over the course of the rotation, please see Section 5 of the SI.

A very similar event happens in aza-butadiene, only now it is the HOMO-1 and HOMO-2 that are anticipated to cross. Generally, aza-butadiene is a much more complicated case and does not lend itself to a simple molecular orbital description as was done with butadiene.[65] Nonetheless, the EHM-ML model remains in quantitative agreement with the underlying DFT throughout the torsional rotation.

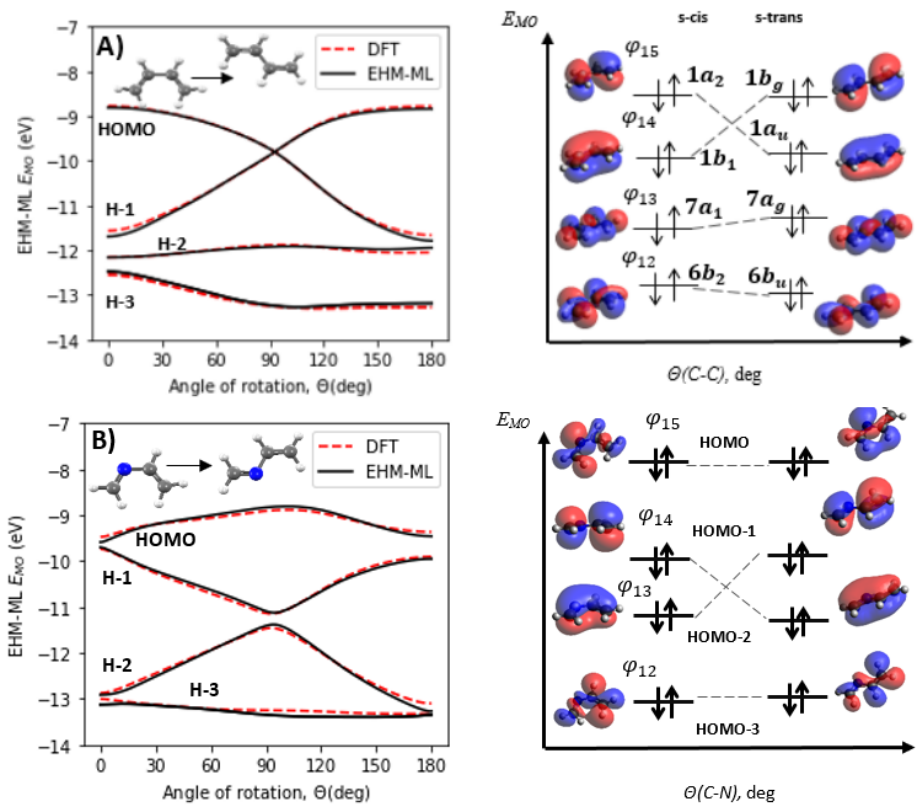

Figure 6: EHM-ML FMO energies (right) and partial FMO diagram (left) which illustrates the orbital changes in the process of internal rotation around central single bond for 1,3-butadiene (A) and 2-aza-1,3-butadiene (B).

## V. CONCLUSIONS

Machine learning is rapidly proving to be a powerful tool for computational chemists. In the past, accurate QM calculations were prohibitively expensive, preventing their application to many interesting problems. By training an ML model to replicate these results, it is now possible to compute high level QM properties at a rate of microseconds per atom. However, results produced from pure ML models are still difficult to interpret. By interfacing ML with a physics-based effective Hamiltonian model for electrons in molecules, namely Hückel theory, we can retain the accuracy of ML, the interpretability of ab-initio calculations, and the speed of a reduced dimensionality description of quantum mechanics. Moreover, mapping the original quantum mechanical problem into a simple model Hamiltonian seemingly provides a viable approach to address the spatial locality challenge for conventional ML schemes targeting individual molecular properties. The later frequently have an empirically determined spatial cut-off radius of 3-6Å[24,25,35,50,66], and may utilize physics models (e.g. electrostatics) to extend beyond this range[67,68]. In contrast, such interactions are naturally introduced through physically transparent terms of the Hamiltonian.

The EHM-ML model provides many surprising and promising results. First, the ML model recovered many of the original parameters used in the EHM. Examples of this include the hydrogen s-orbital energy, as well as the empirical $K^{\ddagger}$ parameter used to modify the off-diagonal elements of the EHM Hamiltonian. Additionally, the EHM-ML model recovered many of the orbital signatures seen in both the h-vibration on methane and the bond rotation on butadiene and aza-butadiene. This was accomplished with HIP-NN rapidly reparametrizing the extended Hückel Hamiltonian as the molecule passed through regions of orbital crossings. Taken in whole, these features strongly point to the EHM-ML model learning and utilizing the underlying physics represented in the EHM to reproduce the quantum effects observed in the electrons of these systems.

The largest shortcoming of this model is the lack of explicit coulomb and exchange terms in the underlying EHM. Neglecting these terms significantly simplified training, as electron density did not need to be iterated to self-consistency. However, their absence also implied that total energy for the EHM-ML model is simply a sum of orbital energies. This is not true for DFT, which prevented multi-objective training to both total molecular energy and valence orbital energy.

Future work should focus on interfacing ML models with more complex effective Hamiltonian representations of Quantum Mechanics, as well as targeting more QM properties in multi-objective training. More complex effective Hamiltonian models (AM1, PM series, MOx, ect.) will facilitate training to total molecular energy, valence orbital energy, and molecular orbital density simultaneously. Possibly, even excitation energies could become a target for training. These more complicated models will also facilitate accurate training to valence orbitals, something that not fully accomplished here without partly destroying the physics encoded in the model. This all points to a very promising future for ML enhanced physics models in quantum chemistry.

## ACKNOWLEDGMENTS

This work was done in part at the Center for Nonlinear Studies (CNLS) and the Center for Integrated Nanotechnologies (CINT) at Los Alamos National Laboratory (LANL). Funding for this work was provided for by the LANL Laboratory Directed Research and Development (LDRD) program. We also acknowledge the LANL Institutional Computing (IC) program and ACL data team for providing computational resources. J.S.S. thanks the Advanced Simulation and Computing (ASC) program for the Nicholas C. Metropolis Postdoctoral Fellowship. O.I. thanks CNLS for their support and hospitality. O.I. acknowledges support from DOD-ONR (N00014-16-1-2311), NSF CHE-1802789. O.I. and T.Z. acknowledge support from NSF EPSCoR RII Grant No. OIA-1632899

# Supplementary Material for "Machine Learned Hückel Theory: Interfacing Physics and Deep Neural Networks"

1. Relative Sizes of Molecules in Data Sets

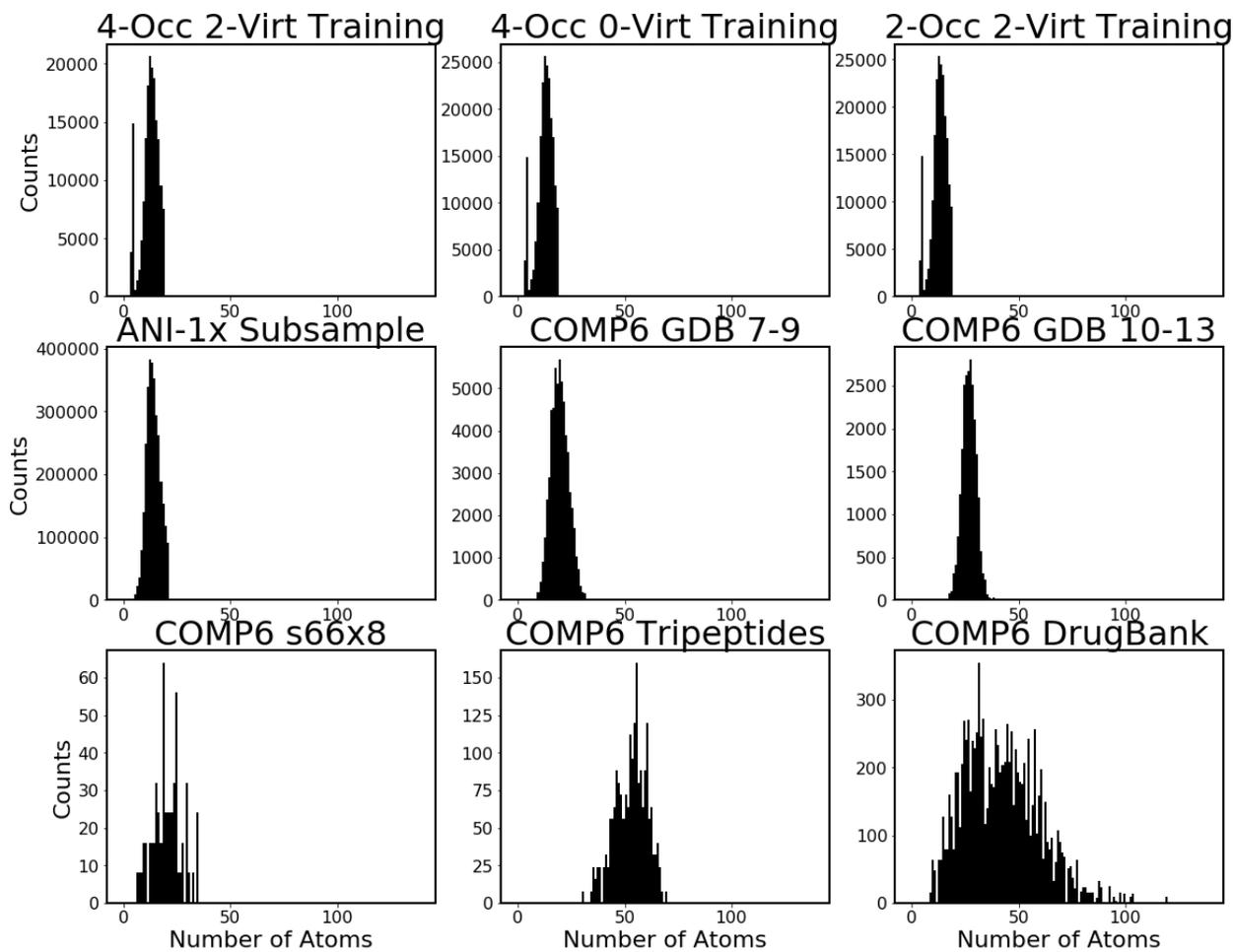

Figure S1: Histograms depicting the sizes of molecules found in training and testing sets used throughout this paper.

Fig. S1 shows the relative molecular sizes of systems found in the training and testing datasets. We would like to emphasize that while the GDB 7-9 and GDB 10-13 subsets are very similar to the underlying training data, the Tripeptides and DrugBank datasets are significantly larger. The success in replicating orbital energies and densities on the Tripeptides dataset shows extensibility.

2. HIP-NN Hückel parameter generation for COMP-6

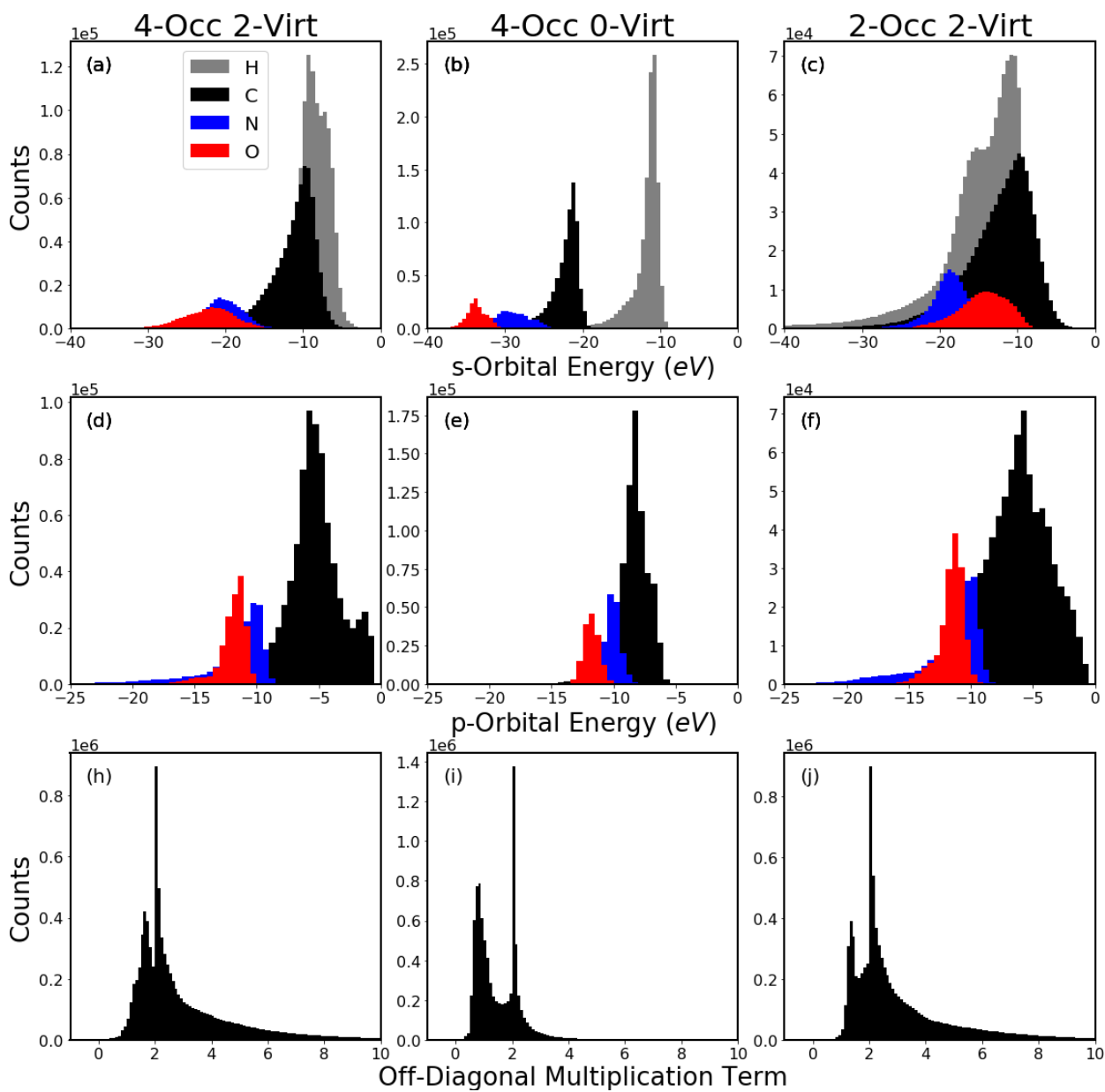

Figure S2: Histograms of ML predicted Hückel theory parameters. The hydrogen s-orbital energy (a,b,c) is in remarkable agreement with the original Hückel theory assignment of -13.6 eV even though the ML model was given no artificial bias to this value. Additionally, though the off diagonal factor is bifurcated in the ML predictions (h,i,j), the average value is in very good agreement with the original parameter used in extended Hückel theory of 1.75.

Fig. S2 shows histograms of the parameterizations generated by HIPNN when predicting on COMP-6. Of the 3 trained models, This figure makes it clear that the "4-Occ 0-Virt" model has the clearest distinction between atom types, with each atom type in Fig. 4b having a clearly distinct distribution. The incorporation of virtual orbitals into the training procedure makes the distributions of parameter predictions indistinct, with all atom types having s-orbital energies over roughly the same range (Fig. 4c). One critical point of agreement between the neural network and the original Hückel theory is the prediction of the hydrogen s-orbital energy. In extended Hückel theory, this parameter was takes from the ionization energy of Hydrogen which is 13.6 eV. By training the neural network on orbital energies, the "4-Occ 0-Virt" network produces an average hydrogen s-orbital 12.5 eV, which is very close to the extended Hückel theory value. Additionally, the distribution of off-diagonal matrix elements (denoted K in fig. 1) is in good agreement with the empirical value of 1.75 originally used in extended Hückel theory. While there are two maxima in the distribution of predicted K values, these maxima do not correlate with atom types in the pair to which the K value is assigned. The reason for these two maxima requires further investigation.

   3. COMP-6 density and gap predictions

Fig. S3 shows the OOF predictions of the three models on the five computed subsections of the COMP-6 benchmark dataset. The models perform about as well on the two GDB sets as they do on the ANI-1x held out validation set. The models perform significantly worse the s66x8 dataset likely due to unusual bonding configurations between molecules not sufficiently covered by the training set. In addition, there is a drop in performance on the Tripeptides dataset, which is likely due to molecular size; as more possible sites (orbitals) are added to a molecule, correctly assigning density to those sites for a specific orbital becomes a more difficult problem.

Fig. S4 shows the HOMO-LUMO energy gap predictions for the three models on the COMP-6 benchmark. The performance of these models at predicting the band gap reflects the performance observed elsewhere.  Predictably, the "4-Occ 0-Virt" model, which was not trained to replicate any virtual orbital properties, fails to predict the band gap entirely. One interesting phenomena is that, while the "4-Occ 2-Virt" model generally has slightly worse predictions on the two GDB sets and s66x8 versus the "2-Occ 2-Virt" model, it does significantly better on the Tripeptides dataset. This may indicate that adding more orbitals to the training set allows the Hückel model to capture more physics and thus generalize to new data better. More investigation into this phenomena is required.

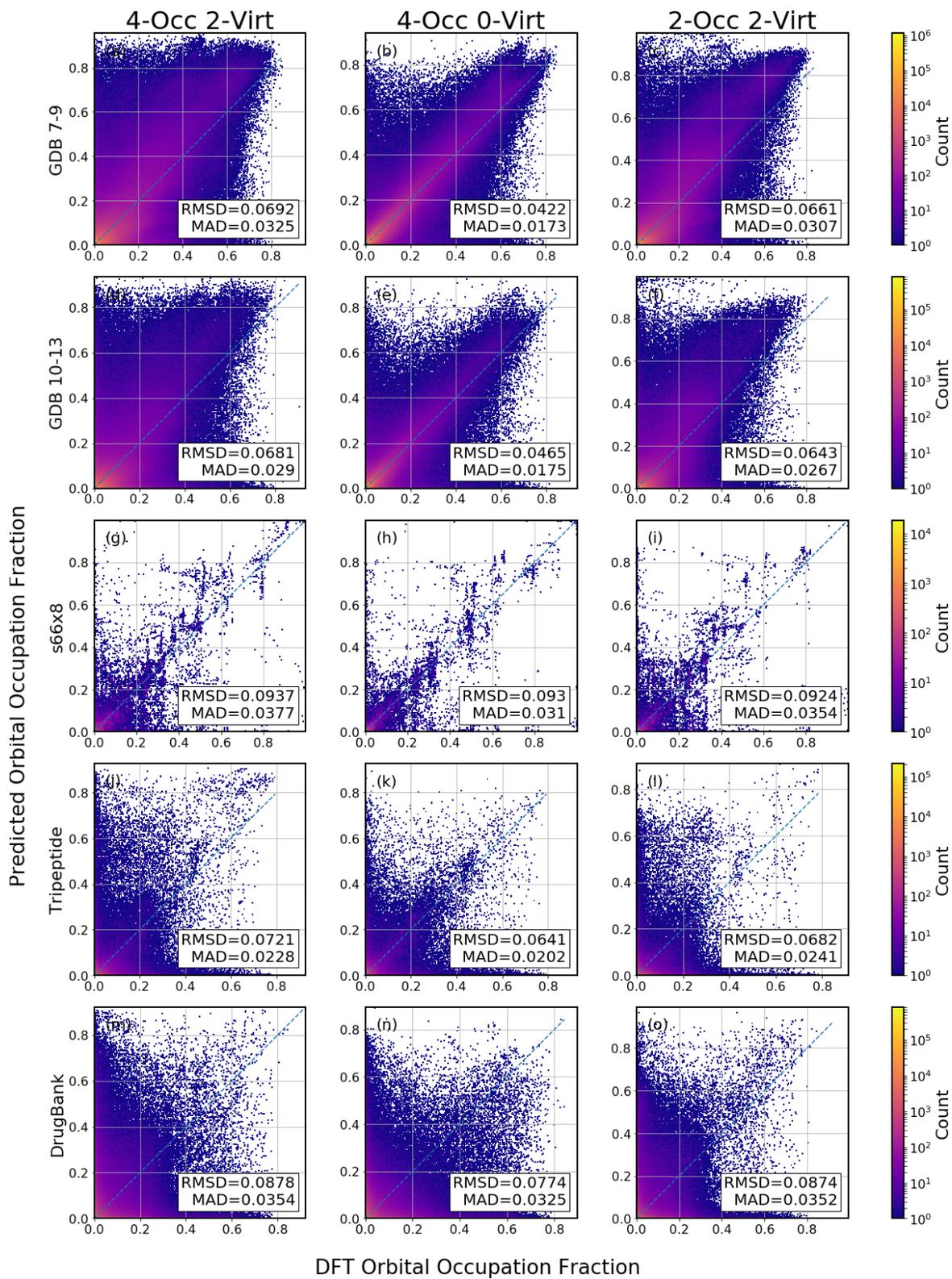

Figure S3: Density predictions of the three models on the 5 computed COMP-6 subsets.

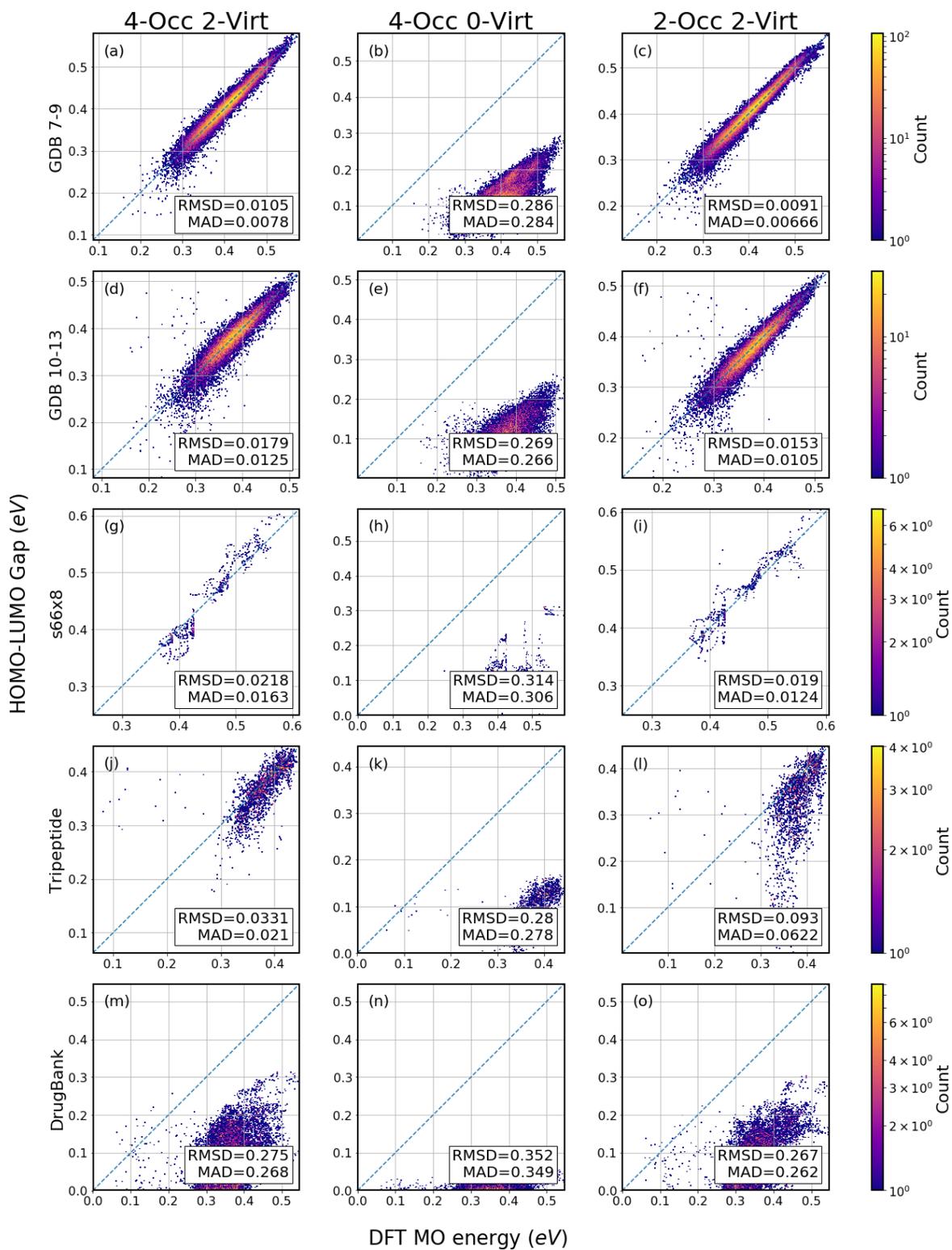

Figure S4: HOMO-LUMO gap predictions of the three models on the 6 computed COMP-6 subsets. Since the "4-Occ 0-Virt" model was not trained on virtual orbitals, it is unsurprising that it fails at HOMO-LUMO gap predictions.

4. Outlier Structures

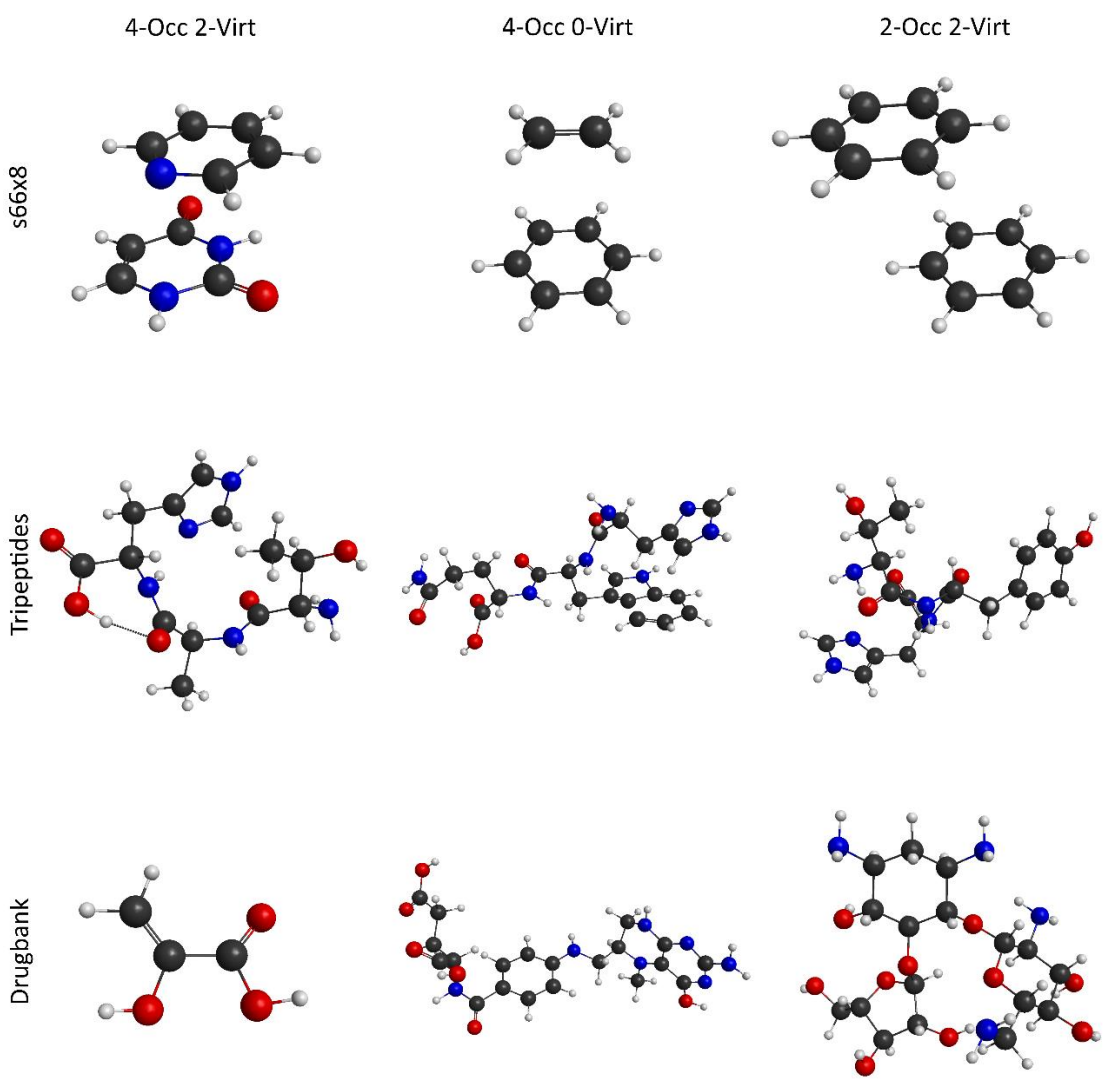

Figure S5: Poorest performing molecular structures by network and dataset for COMP-6. The GDB sets were not included due to the consistently high accuracy of the networks on those datasets.

In Fig. S5 we show the worst performing structure by network and COMP-6 dataset. With one exception, every structure has a ring system and such structures are known to be under sampled in the training set. Future work should focus on augmenting the training set with more organic ring structures, specifically aromatic and anti-aromatic structures.

5. Butadiene predicted orbital parameters.

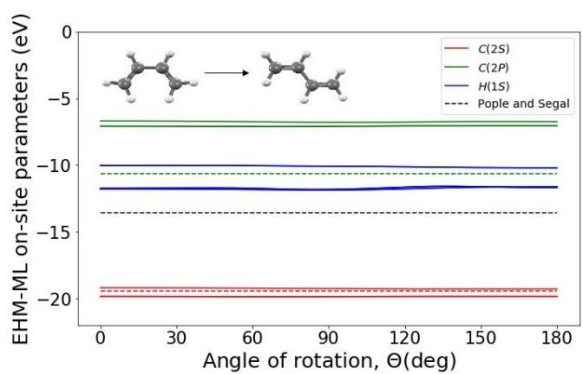

Figure S6: Atomic site parameters for butadiene as it undergoes internal rotation.

Fig. S6 shows the atomic site parameters for butadiene as the molecule undergoes internal rotation. Interestingly, unlike in the methane case, these parameters are very nearly constant. This is likely due to rotations not representing significant changes in molecular environment according to HIP-NN. Despite this fact, the Hückel models still shows quantitative accuracy when predicting the 4 highest lying occupied orbitals of butadiene, further indicating that much of the physics occurring in these molecules is captured by the Hückel model.